\RequirePackage{ifpdf}
\documentclass{PoS}
\usepackage[sort,round]{natbib}
\usepackage{float}
\usepackage{epsfig}
\usepackage{graphicx}
\usepackage{graphics}
\usepackage[latin1]{inputenc}
\usepackage{latexsym}

\def\gsim{\mathrel{\rlap{\lower 4pt \hbox{\hskip 1pt $\sim$}}\raise 1pt
\hbox {$>$}}}
\def\lsim{\mathrel{\rlap{\lower 4pt \hbox{\hskip 1pt $\sim$}}\raise 1pt
\hbox {$<$}}}

\title{Investigations of supernovae and supernova remnants in the era of SKA}

\ShortTitle{SNe and SNRs with SKA}

\author{Lingzhi Wang, Xiaohong Cui, Hui Zhu, Wenwu Tian \\
        National Astronomical Observatory of China, Chinese Academy of Sciences, Beijing 100012, China\\
        E-mail: \email{wanglingzhi@bao.ac.cn}, \email{tww@bao.ac.cn}} 
 
 \author{Xiaofeng Wang\\
        Department of Physics, Tsinghua University, Beijing 100084, China\\}
        
\abstract{Supernovae are extremely luminous and can outshine an entire
galaxy for a period of days. Two main physical mechanisms are used to
explain supernova explosions: thermonuclear explosion of a white dwarf
(Type Ia) and core collapse of a massive star (Type II and Type Ib/Ic).
Type Ia supernovae serve as distance indicators that led to the discovery
of the accelerating expansion of the Universe. The exact nature of their
progenitor systems however remain unclear. Radio emission from the
interaction between the explosion shock front and its surrounding
circumstellar medium (CSM) or interstellar medium (ISM) provides an
important probe into the progenitor star's last evolutionary stage. No
radio emission has yet been detected from Type Ia supernovae by current
telescopes. The SKA will hopefully detect radio emission from Type Ia
supernovae due to its much better sensitivity and resolution.

There is a 'supernovae rate problem' for the core collapse supernovae
because the optically dim ones are missed due to being intrinsically faint
and/or due to dust obscuration. A number of dust-enshrouded optically
hidden supernovae should be discovered via SKA1-MID/survey, especially for
those located in the innermost regions of their host galaxies. Meanwhile,
the detection of intrinsically dim SNe will also benefit from SKA1. The
detection rate will provide unique information about the current star
formation rate and the initial mass function. 

A supernova explosion triggers a shock wave which expels and heats the
surrounding CSM and ISM, and forms a supernova remnant (SNR). It is
expected that more SNRs will be discovered  by the SKA. This may decrease
the discrepancy between the expected and observed numbers of SNRs. Several
SNRs have been confirmed to accelerate protons, the main component of
cosmic rays, to very high energy by their shocks. This brings us hope of
solving the Galactic cosmic ray origin's puzzle by combining the low
frequency (SKA) and very high frequency (Cherenkov Telescope Array: CTA)
bands' observations of SNRs.  }

\FullConference{
Advancing Astrophysics with the Square Kilometre Array\\
June 8-13, 2014\\
Giardini Naxos, Italy}

\begin{document}
% If the speaker is not the first author, then enter the first author
% information below, so it appears in the header of each page
% this must go AFTER the \begin{document} command
\makeatletter
\setbox\@firstaubox\hbox{\small Lingzhi Wang}
\makeatother

\section {Introduction}
The SKA will be a giant radio imaging and survey telescope array that will have an effective collecting area of one-million square meters and will be operated in the wide band range between 70~MHz to 10~GHz. The combination of its unprecedented sensitivity and resolution will start a new era of exploring the electromagnetic and other windows into the Universe. One of its key sciences (radio transients) will involve looking at vast swathes of the radio sky and making a detailed map as the thousands of telescopes work together in unison. Radio surveys have already recorded many transient events which exhibit rapid changes and large amounts of energy over short periods of time, e.g., supernovae (SNe) and their SN remnants (SNRs). With its higher sensitivity and broader field of view, the SKA will bring a revolution in the field of SNe and SNRs.

The low-frequency part (SKA1-LOW \& SKA1-SUR)  of SKA will be built in Western Australia where it will be merged with its precursor ASKAP, while the mid- and high-frequency part (SKA1-MID) of SKA will be built in South Africa where it will be merged with its precursor MeerKAT. Table \ref{phase1} shows the parameters of SKA phase 1 and its precursors, and also includes parameters for the Cherenkov Telescope Array (CTA) for comparison.

\begin{table}[h]
  \caption{Parameters for SKA phase 1 telescopes (SKA1-LOW, SKA1-MID, \& SKA1-SUR) and their precursors (ASKAP \& MeerKAT), together with the Cherenkov Telescope Array.}
\begin{center}
 \begin{tabular}{cccccc}
  \hline
Parameter&Resolution$^a$ &FoV$^a$ &Bandwidth &Sensitivity\\
         &(arcsec) &(deg$^2$) &(MHz) &($\mu$Jy hr$^{-1/2}$)\\
\hline
SKA1-LOW &11  &$\sim$30  &250  &$\sim$2 \\
SKA1-MID &0.22&$\sim$0.5 &770  &$\sim$1 \\
SKA1-SUR &0.9 &18  &500  &$\sim$4 \\
ASKAP    &7   &30        &300  &$\sim$30\\
MeerKAT  &11  &0.86      &1000  &$\sim$3\\
\hline
CTA     &0.04-0.05~deg & 6-8~deg     &     &10$^{-13}{\rm erg~cm^{-2}~s^{-1}}$ \\
\hline 
\end{tabular}
\end{center}
\vspace*{-3pt}
$^a$Fiducial Frequencies assumed: 110 MHz for SKA1-LOW; 1.4 GHz for SKA1-MID
and SKA1-SUR; 1.4 GHz for ASKAP and MeerKAT \citep{dewdney13}; 1 TeV for
CTA.
  \label{phase1}
  \end{table}

The SKA offers an excellent opportunity to systematically study the radio emissions from SNe and SNRs thanks to its  higher sensitivity and wide FOV. The paper is organized as follows: \S2 presents the radio SNe and their applications; \S3 is the research of SNRs and cosmic rays; \S4 draws our conclusions.

\section {Radio SNe Investigations in the era of SKA1}
Two main physical mechanisms are used to explain the SNe observations. One is thermonuclear explosion of a white dwarf in a binary system i.e., Type Ia SNe; the other is core collapse of a massive star ($M_{MS} \gsim 8~M_{\odot}$), e.g., Type II, Ib/c SNe and their subtypes \citep{Filippenko97,Smith14}. SNe play a critical role in galactic evolution, via explosive nucleosynthesis and chemical enrichment, energy injection into the CSM \& ISM, long $\gamma$-ray bursts and the origin and acceleration of cosmic rays. Moreover, Type Ia SNe are used as cosmological distance indicators due to their nearly homogeneous intrinsic luminosities. Understanding the progenitor stars and explosion mechanisms for the different SNe types are the primary goal of SN investigation, which spans optical, X-rays, and radio observations. Radio emission from SNe involves the acceleration of relativistic electrons and amplified magnetic field necessary for synchrotron radiation, producing from the SN shock interactions with a relatively high-density CSM, which is assumed to be formed by the mass loss in the late stages of the progenitor's evolution. Intriguingly, the mass loss rate can potentially be used to probe different types of SN progenitor stars, see Table \ref{progenitor} \citep{Smith14}. Therefore, radio emission from SNe plays a critical role in studying their progenitors' properties and final stages of evolution.

Radio SNe have been extensively studied for decades and detected at multiple radio frequencies with facilities including the Very Large Array \citep[VLA;][]{Weiler07}. The radio light curves can be fit by the parameterized model developed by \cite{Weiler86,Weiler90,Weiler02}, which can be used to compute the ratio of the mass loss rate to the stellar wind velocity. Different SN progenitor stars have their own characteristic mass loss rates (Table \ref{progenitor}), shaping different CSM environments and thus leading to a diversity of radio light curves \citep{Weiler02,Wellons12}. Radio flux modulations are likely attributed to the CSM density modulations, which can be used to trace the mass loss rate variations of pre-SNe late evolutionary stages, and the influence of binaries. This method has been applied to Type II \citep{Weiler02} and Ibc SNe \citep{Wellons12} to constrain SNe CSM environments.

\begin{table}[ht]
  \begin{center}
\caption{Mapping of SN types to their likely progenitor star properties. Taken from \cite{Smith14}}
\label{progenitor}
\begin{tabular}{llccc}
  \hline
  SN       &Progenitor Star  &$M_{ZAMS}$     &$\dot{M}$ &$V_{wind}$     \\
  ...      &...         &($M_{\odot}$)  &($M_{\odot}$ yr$^{-1}$) &(km s$^{-1}$)   \\ 
  \hline
  II-P     &RSG            &8--20         &10$^{-6}$--10$^{-5}$    &10-20        \\
  II-L     &RSG/YSG        &20--30 (?)    &10$^{-5}$--10$^{-4}$    &20-40        \\
  II-pec   &BSG (b)        &15--25        &10$^{-6}$--10$^{-4}$    &100-300        \\
  IIb      &YSG (b)        &10--25        &10$^{-5}$--10$^{-4}$    &20-100       \\
  Ib       &He star (b)    &15--25 (?)    &10$^{-7}$--10$^{-4}$    &100-1000     \\
  Ic       &He star (b)/WR &25--?         &10$^{-7}$--10$^{-4}$    &1000         \\
  Ic-BL    &He star (b)/WR &25--?         &10$^{-6}$--10$^{-5}$    &1000         \\
  \hline
  IIn (SL) &LBV            &30--?         &(1--10)                &50-600      \\
  IIn      &LBV/B[e] (b)   &25--?         &(0.01-1)               &50-600      \\
  IIn      &RSG/YHG        &25--40        &10$^{-4}$--10$^{-3}$    &30-100       \\
  IIn-P    &super-AGB      &8--10         &0.01-1                 &10-600       \\
  Ibn      &WR/LBV         &40--?         &10$^{-3}$--0.1          &1000        \\
  \hline
  Ia-CSM   &WD (b)         &5-8 (?)       &0.01-1                 &50-100      \\
 
  \hline
\end{tabular}
\end{center}
RSG: Red supergiant; YSG: Yellow supergiant; BSG: Blue supergiant; WR: Wolf-Rayet star; \\
LBV: Luminous blue variable; B[e]:  B-type stars; YHG: Yellow hypergiant; WD: white dwarf \\
(b) indicates probably a binary channel.
  \end{table}

\subsection {Progenitors of Type Ia SNe via SKA1}
Two scenarios of single degenerate (SD) and double degenerate (DD) progenitor systems dominate the current SNe explosion models. The post-explosion radio detection of the blastwave impact with the CSM would not only favor the SD model, but also provide the density and structure of the CSM. Due to extremely low mass loss rates, no radio emission has yet been detected from Type Ia SNe \citep{Panagia06,Hancock11}, even for the closest SNe 2011fe \citep{Chomiuk12} and 2014J \citep{Perez14}, which have set tight upper limits on radio luminosity, hence placing some constraints on the mass loss rate and wind velocity from its progenitor system. The lack of a radio detection from either SN 2011fe (d $\sim$ 6.4~Mpc) or SN 2014J (d $\sim$ 3.5~Mpc) all but rules out a symbiotic progenitor system and also a system with high accretion rate onto the white dwarf \citep{Chomiuk12,Perez14}.

\begin{figure}[ht]
  \begin{center}
\includegraphics[width=0.7\textwidth]{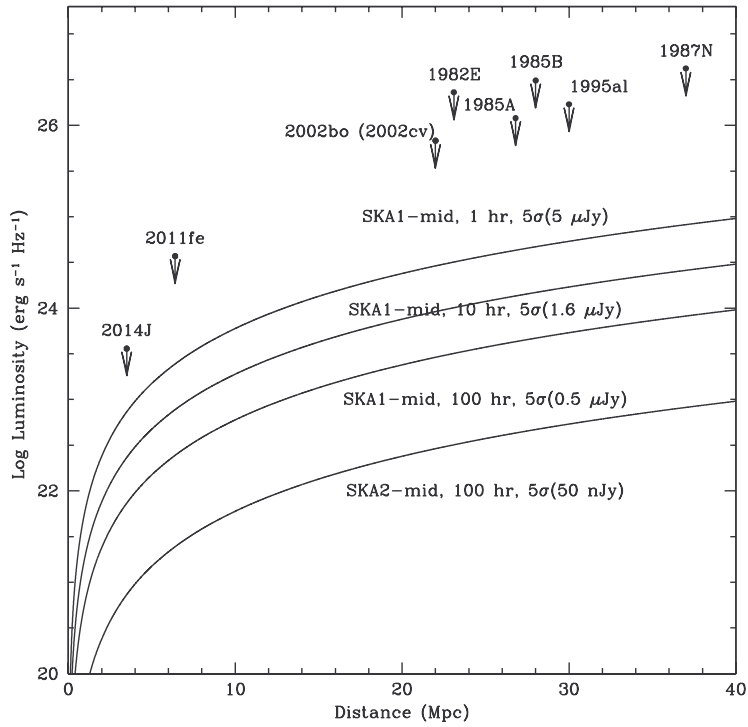}
\caption {Log luminosity VS distance plot for Type Ia SNe by SKA1-MID in 1, 10, 100 hr and SKA2-mid in 100 hr observations. Eight close Ia SNe were indicated by downward arrows. 
 }
\label{typeia}
\end{center}
\end{figure}

Due to their intrinsic faintness in the radio and their low volumetric rate, blind detection of Type Ia SNe is unfeasible \citep{Panagia06,Lien11}. SKA1-MID will hopefully carry out targeted follow-up observations to detect the first radio light from nearby Type Ia SNe ($\lsim$20~Mpc) due to its much better sensitivity and resolution (Figure \ref{typeia}). Otherwise, no radio detection from Ia SN will place stronger constraint on its last evolutionary stage of progenitor system (e.g. lower mass loss rate upper limit). Since the sensitivy of SKA1-MID will be improved to 10 $\times$ higher ($0.1{\rm \mu Jy~hr^{-1/2}}$) than the current one at phase 2, radio luminosity from a Type Ia SN could be detected as low as $2.4\times10^{22}{\rm ~erg~s^{-1}Hz^{-1}}$, out to the Virgo Cluster ($\sim$20~Mpc) after 100~hr integrations as a 5$\sigma$ detection, as shown by the bottom solid line from Figure \ref{typeia}.

\subsection {CCSNe survey}

CCSNe represent the core collapse death of massive stars.  The cosmic CCSNe rate can be determined by multiplying cosmic star formation rate (SFR) by the efficiency of forming CCSNe. \citet{Horiuchi11} raised the "SN rate problem": the measured cosmic CCSNe rate is a factor of $\sim$2 smaller than that predicted from the cosmic massive-star formation rate. The "missing" percentage of optical CCSNe surveys can reach up to $\sim$30\% at $Z=1$ and increase to $\sim$60\% at $Z=2$ \citep{Mannucci07,Mattila12}. Many CCSNe are missed due to being intrinsically dim or due to obscuration. Radio techniques have great potential to find more "missing" CCSNe, especially CCSNe hidden in the inner regions of their host galaxies, because radio photons are not influenced by dust. SKA1-SUR provides a good chance to discover more faint "missing" CCSNe to yield a better sample because of its higher sensitivity and larger field (Table \ref{phase1}). 

\begin{figure}[ht]
  \begin{center}
\includegraphics[width=0.7\textwidth]{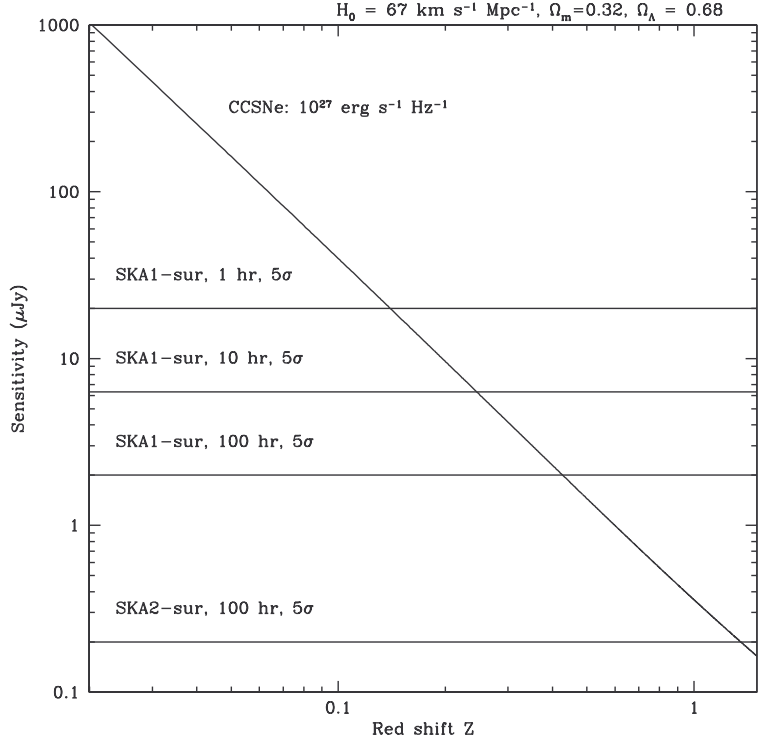}
\caption {Sensitivity VS redshift plot for CCSNe with typical average radio luminosity of $10^{27}~{\rm erg~s^{-1}~Hz^{-1}}$ at 1.67~GHz. 
 }
\label{ccsne}
\end{center}
\end{figure}

Typical radio luminosities of CCSNe are $10^{27}~{\rm erg~s^{-1}~Hz^{-1}}$ \citep{Lien11, Kamble14}. Luminosity distance is $D_L(z) = (1+z) \ c/H_0 \ \int^z_0 dz^\prime \ [\Omega_{\rm m} (1+z^\prime)^3 + \Omega_{\Lambda}]^{-1/2}$, where the cosmological parameters were adopted from \citet{Plank13}. SKA1-SUR will reach to Z = 0.14 (225~Mpc) in one hour, and to Z = 0.25 (0.43) in 10 (100) hr observations, refer to Figure \ref{ccsne}. For a more complete outline of a CCSNe survey, also refer to \citet{perez14b}.

\subsection  {SNe and SNRs connections}
The transition from SNe to SNRs, or young SNRs to old SNRs are very critical for understanding SNe evolution and their interactions with CSM and ISM. Currently there are gaps between young SNRs and old SNRs. Bridging this gap will help to understand the spectra index variation and the magnetic field direction variation in SNRs. For example, the magnetic field in young SNR Tycho shells is predominantly radially oriented, while that in evolved SNR CTB1 shells is largely tangential to the shock front \citep{Furst04}.

\section {SNRs investigations in the era of SKA}
After SNe exploding, the results of interaction between ejected steller material and the ambient gas, form SNRs, refer to the review \citet{Woltjer1972}. Four-zone structure of a SNR are constituted with the freely expanding ejecta, the shocked ejecta, the shocked ambient medium (AM including CSM \& ISM) and the unperturbed ambient gas. The SNRs have four evolutional stages, see Table \ref{evolution}. More mass the exploding ejecta swept up, slower shock velocity expanded and bigger size the SN remnants evolved. During the evolution process, the ejectors inject the huge energy into the AM. The feedback machanism of the exploding energy ($\sim10^{51}$~erg) and matter leads to metals richening and heating in the AM. SNRs represent the last evolution stage of massive stars, which can be taken as agents of testing massive stellar evolution theory.

\begin{table}[ht]
  \caption{The evolution phases of SN remnants}
  \begin{tabular}{crrrrl}
    \hline
   & Mass swept up & Velocity     & Radius   & Age          &Phase\\
   & M$_{\odot}$    & kms$^{-1}$    & pc       & yrs          & \\
    \hline
 I & <5            & <10,000      & <1       &<2,000        &ejecta-dominated (free expansion)\\
 II& tens          &$\sim$ 200    &$\sim$ 10 &$\sim$40,000  &Sedov-Taylor(adiabatic)\\
III&$\sim$ 1,000   &$\sim$ 20     &$\sim$ 30 &$\sim$100,000 &pressure-driven-snow-plow\\
IV &               &              &          &              &momentum-conserving (merging)\\
\hline
\end{tabular}
  \label{evolution}
  \end{table}

\subsection {SNRs' discovery}
SNRs play a key role in massive stellar evolution, thus the number of SNRs in a galaxy can be used as a diagnostic tool to test the current star formation. Taking our Galaxy for example, about 300 SNRs have been detected. Most of them were discovered by radio surveys. \citep{Green09a,Green09b}. However, the predicted number of SNRs is more than one thousand \citep{Li91}. The reason existing the big gap is likely due to selection effects acting against the discovery of old, faint, large remnants ($<$0.3~Jy), as well as young, small remnants ($<$1~arcmin) in previous limited-sensitivity and/or -resolution Galactic radio surveys. In addition, the missing SNRs are likely concentrated toward the inner Galaxy, where thermal HII regions cause the most confusion due to their shell-like morphologies. The previous methods of seeking SNRs are based on their radio continuum data from high frequency and low sensitivity and skews the statistics, therefore only the bright and big SNRs can be detected (Green 2009). More sensitive, higher resolution surveys of the inner Galaxy at low radio frequencies are perhaps the key to determine how many SNRs are missing or to test current predicted SNRs rates.

SKA has a sensitivity of A$_{\rm eff}$/T$_{\rm sys}$=10000 and resolution of less than 1~arcsec. The fiducial sensitivity can reach as low as 50~nJy with 100 hours exposure time at 1.4~GHz. The radio emissions of SNRs are nonthermal and follow power law flux density as a function of frequency with a typical spectral index -0.5. The SKA1-LOW, with frequency range of between 50~MHz to 350~MHz, hosts great potentials to discover a large amount of 'missing' SNRs in Galaxy at the sensitivity of $\sim$2~$\mu$Jy~hr$^{-1/2}$ with a FOV of $\sim$30 deg$^2$, see Table \ref{phase1}.

\subsection {High energy Cosmic rays origin via SKA \& CTA}
The work of identifying two SNRs as Cosmic Rays (CRs)' accelerators was selected as the Breakthough of the Year 2013 by the magazine <Science>. <Science> claimed that astronomers traced high-energy particles called cosmic rays back to their birthplaces in the debris clouds of SNe \citep{Ackermann13}. The research in complex particle acceleration in an SNR is undergoing notable growth recently. The multi-wavelength observations of SNRs, especially the high-energy $\gamma$-rays observations from ground-based and space-based telescopes, play a powerful tools to probe the particle acceleration machanism in SNRs. Shell-type SNRs (e.g. young: G347.3-0.5; see \citet{Enomoto02} and old: G353.6-0.7, see \citet{Tian08}) are the major component of the five Galactic populations to generate very high energy $\gamma$-rays, i.e., pulsar wind nebulae (PWNe), X-ray binaries, shell SNRs, young stellar cloud, and giant molecular cloud.

\begin{figure}[h]
  \begin{center}
\includegraphics[width=.8\textwidth]{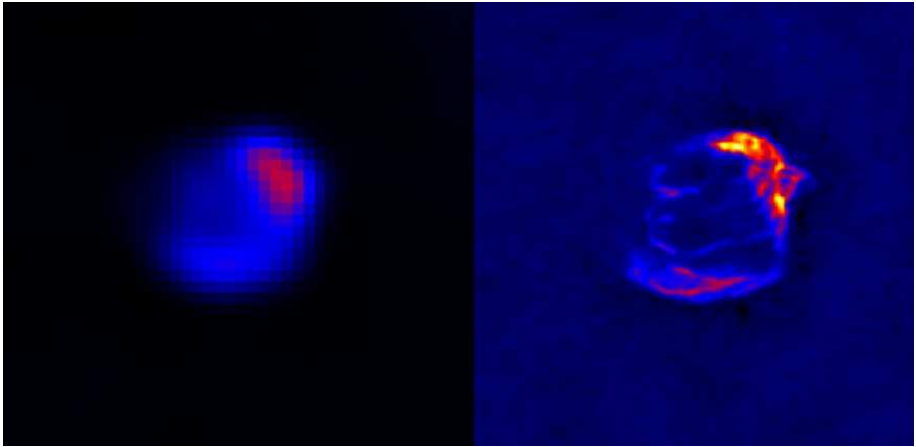}
\caption{1.4~GHz continuum observations of SNR W49B with two resolutions: Left, $\sim$1~arcmin \citep{VGPS}; right, $\sim$6~arcsec \citep{MAGPIS}. Filaments are seen in right panel, while not in left panel. }
\label{resolution}
\end{center}
\end{figure}

Magnetic field amplification is needed to accelerate cosmic rays to very high energy in SNRs. Meanwhile, the density perturbations come up and are most rapid for the smaller scales or filaments, when passing through the shock. Observations of filaments of non-thermal X-ray emission in all young SNRs provide strong evidences of magnetic field amplification in the shock regions \citep{Vink12}. Therefore, filamentary emissions are very important to study the particle acceleration. Taking figure \ref{resolution} for example, only higher resolutions telescopes can map the filaments' details. Theoretical calculations and simulations reveal that the amplification usually happens on the scale of $<$0.1~pc. SKA will host great potentials to observe these structures due to its high resolution of less than 1~arcsec which could distinguish the structures of about 0.02~pc at a distance of 5~kpc and reduce the depolarization caused by small scale distortion of magnetic field. 

Cosmic rays are mainly composed of charged particles which are deflected in the magnetic fields, thus, direct observations on the charged particles carry little information on their origin. However, observations of $\gamma$-ray photons, as small component of the CRs ($<$1\%) play a key role in solving this issue due to two reasons: $\gamma$-rays have straight line propagation; $\gamma$-rays may be produced by relativistic particles' interactions \citep{Tian13}. Theoretical exploration has indicated two major emission mechanisms for $\gamma$-rays from Galactic objects: hadrons and leptons origin. When accelerated protons interact with interstellar material they generate neutral pions, which in turn decay into $\gamma$-rays (hadron origin). However, the high-energy electrons also generate $\gamma$-rays by means of inverse Compton scattering and bremsstrahlung. 

The low-frequency synchrotron emission ($\lsim$2~GHz) tracks the Mev-Gev part of electrons' spectra. For example, supposing a typical Galactic magnetic field of 10~$\mu$G, electrons with 2.9 GeV energy will have their peak radio emission at a frequency 408~MHz; electrons of 5.4 GeV energy will have peak emission at 1420~MHz; and electrons of 7.5 GeV energy will have peak emission at 2695~MHz \citep{Leahy09}. The relativistic electrons are generally understood to be accelerated by the Fermi mechanism in fast moving shock fronts. There is strong evidence of electron acceleration based on the detection of synchrotron X-rays from shell of young SNRs \citep{Koyama95}. For old SNRs with an age of around 10$^5$~yr, theoretical calculations show that primary electrons could not be accelerated to high energy enough to emit obvious $\gamma$ rays. But proton acceleration is still efficient to produce $\gamma$-ray emission. In lower frequency, non-thermal emissions from the secondary electrons generated in $pp$ collisions dominate the spectrum. The energy spectrum of such secondary electrons starts to deviate from a pure power below $\sim$1~GeV and finally to cut off below $\sim$100~MeV. In turn, the radio continuum synchrotron spectrum from the secondary electrons will start to diverge from the pure power at frequencies below 100~MHz in case of typical B of 10~$\mu$G \citep{Aharonian13}. Therefore, The SKA1-LOW with frequency range of between 50~MHz to 350~MHz, provides an excellent opportunity to distinguish the spectrum features of old SNRs produced from protons or secondary electrons. Obviously the low frequency radio spectrum serves as the proxy of the GeV energy particles. For high energy TeV particles, X-rays synchrotron spectrum ($~$KeV) features are a powerful diagnostic tool to probe the TeV particles. In the coming golden era, observation of both sensitive $\gamma$-ray and low frequency radio telescopes, i.e., CTA and SKA1-LOW, will simultaneously trace the emissions produced by high energy GeV \& TeV particles from SNRs so help solving the CR's origin issue. In addition, the high sensitivity of SKA is very useful to search for the radio counterpart of unidentified $\gamma$-ray source. 

\section{Summary}

SNe and SNRs play a key role from stellar scale (for stellar evolutions) to galaxy scale (for galaxy chemical evolution), and out to cosmological scale (as distance candles). Radio emission from SNe will probe the last evolutionary stages of their progenitor systems and test the stellar evolutionary theory. The radio and $\gamma$-ray emission from SNRs will hopefully address the cosmic rays' origin. SKA1 will open a new era for investigations of SNe and SNRs, thanks to its higher sensitivity and better resolution.   

\setlength{\bibsep}{0.0pt}
\bibliography{skasn}{}
\bibliographystyle{apj}

\end{document}